\documentclass[%
 reprint,
 amsmath,amssymb,
 aps,
]{revtex4-2}

\usepackage{graphicx}
\usepackage{dcolumn}
\usepackage{bm}

\begin{document}

\preprint{APS/123-QED}

\title{Electron nonlinear dynamics in a compact accelerator based on the circular rotating TM$_{110}$ mode\\}
 
\author{E. A. Orozco}
\email{eaorozco@uis.edu.co}
\author{P. Tsygankov}%
\author{F.F Parada-Becerra}%
\author{A. Hernández}%
\author{Y. Barragán}%
\affiliation{%
 Universidad Industrial de Santander. A.A. 678 Bucaramanga, Colombia.\\
}%
\author{A. D. Martínez}
\homepage{http://www.udes.edu.co/} 
\affiliation{
 Universidad de Santander. Bucaramanga, Colombia.\\
}%


\date{\today}

\begin{abstract}
An electron autoresonant acceleration by the rotating TM$_{110}$ mode microwave field in an inhomogeneous magnetostatic field is studied. A set of differential equations describing the evolution of the phase shift between the electron angular position and the angle for which the transferred power is maximum, the total electron energy, and the longitudinal electron velocity are obtained. Magnetic field profiles to keep the electron in the acceleration regime are found.\\
The results show that an electron injected along the cavity axis with an energy of the $30$ $keV$ can be accelerated up to energies about of $200$ $keV$, using an electric field amplitude of 20 kV/cm, a frequency of 8 GHz and a linear magnetic field profile. Also, we consider the case of electron acceleration in exact resonance conditions. The corresponding magnetic field profile predicted by the model was found\\
The results presented in this paper can be useful for designing RF accelerators, based on the circular rotating $TM_{110}$ mode, used in x-ray sources for medical applications or airport security, among others.  
\end{abstract}

\maketitle


\section{\label{sec:Intro}INTRODUCTION}

The electron cyclotron resonance (ECR) phenomenon occurs when an electron affected by a magnetic field interacts with an electromagnetic wave whose frequency coincides with the electron cyclotron frequency. The self-sustenance of the cyclotron resonance interaction between the electron and the transverse electromagnetic wave which is propagated along the homogeneous magnetic field has been known since 1962  \cite{field1963resonance, davydovski1962possibility}. Said condition is maintained automatically, in spite of the decrease in the cyclotron frequency caused by the increase in electron energy. The condition of the particle-wave cyclotron resonance is written down as \cite{milant2013cyclotron} 
\begin{equation}
\omega-k v_z=\frac{\omega_{\mathrm{c} 0}}{\gamma} \equiv \omega_{\mathrm{c}} \label{Res_condition_1}
\end{equation}
where $\omega$ and $k$ are the frequency and wave number of the electromagnetic wave, respectively; $v_z$ is the particle velocity component along the magnetic field $\mathbf{B}_0$, $\omega_{\mathrm{c} 0}=q B_0 /(m \omega)$ is the classical cyclotron frequency, being $q$ and $m$ the charge and mass of the particle, respectively; $\gamma$ is the Lorentz factor and $\omega_{\mathrm{c}}$ is the cyclotron frequency. The resonance condition (\ref{Res_condition_1}) is preserved because the decrease in the cyclotron frequency is exactly compensated by the Doppler shift. Based on said acceleration mechanism, other studies have been developed  \cite{loeb1986autoresonance,salamin2015feasibility}. Jory and Trivelpiece studied the electron acceleration by considering linearly and circularly polarized, homogeneous plane waves, as well as linearly and circularly polarized $TE_{11}$ modes in waveguides of circular cross-section in a nonuniform steady axial magnetic field \cite{jory1968charged}. R. Shpitalnik et al proposed a novel autoresonance microwave accelerator (AMA), based on a circularly polarized $TE_{11}$ mode, and a spatially tailored magnetostatic field that allows continuous electron acceleration \cite{shpitalnik1991autoresonance}.\\ 
The particle acceleration in cavity resonators of circular cross section ($TE_{111 }$ mode) immersed in a uniform steady axial magnetic field also has been studied \cite{mcdermott1985production}. Due to the energy dependence of the cyclotron frequency, in this acceleration scheme, it is not possible to maintain resonance over the entire trajectory. Therefore, an optimum magnetic field was used, which occurs when the rf frequency is roughly equal to the average cyclotron frequency, i.e., $\omega \approx 2 \omega_{c 0} /\left(1+\gamma_f\right)$, where $\gamma_f$ represents the electron $\gamma$ as it exits the cavity \cite{mcdermott1985production}.\\
ECR conditions can be maintained over time through the GYRAC mechanism proposed by Golovanivsky; which is realized in a single-mode standing microwave field and a homogeneous magnetic field that increases with time to compensate the growth of the relativistic factor \cite{golovanivsky1980autoresonant,golovanivsky1982gyrac}. Based on the GYRAC mechanism, the formation of plasma bunches in a long mirror machine under the gyromagnetic autoresonance was numerically studied in a magnetic field that increased with time \cite{andreev2020generation}.\\
A. Neishtadt and A. Timofeev showed that autoresonance acceleration can play an important role in electron cyclotron heating of the plasma because electrons can keep the ECR condition along a magnetic field line as they move towards regions where the magnetic field increase \cite{neishtadt1987autoresonance}. This kind of autoresonance mechanism can be named spatial autoresonance; which has been used as a basis to design accelerator cavities of electron beams \cite{dugar2009cyclotron,dugar2017compact,otero2019numerical,velazco2003development,velazco2016novel}. In said systems, the selected magnetostatic field profile maintains the equality between the microwave frequency and the cyclotron frequency along the electron trajectory. A spatial autoresonance acceleration (SARA) based on a Transversal electric field,  $TE_{11p}$ $(p=1,2,3)$, was numerically studied \cite{dugar2009cyclotron}. An X-ray source based on the SARA concept has been patented \cite{dugar2017compact}. A novel compact rotating-wave electron beam accelerator (RWA) based on spatial autoresonance and the  Transversal magnetic field, $\mathrm{TM}_{110}$,  was proposed by Velazco et al \cite{velazco2003development, velazco2016novel}. \\
In the present paper, a theoretical model is described for studying electron autoresonant acceleration by the rotating circular TM$_{110}$ mode in an inhomogeneous magnetostatic field, using a single-particle approximation and considering the nonlinear regime. Equations are derived for calculating both the electron cyclotron frequency and the Larmor radius. A set of differential equations is obtained to describe the evolution of the phase shift between the electron angular position and the angle for which the transferred power is maximum, as well as the variations in total energy and longitudinal velocity, which are solved numerically. The results obtained in this work show good agreement with those obtained from the numerical solution of the relativistic Newton-Lorentz equation by using the Boris method \cite{birdsall2004plasma, qin2013boris}.

\section{\label{sec:Formalism}THEORETICAL FORMALISM}

\subsection{\label{sec:Physchem}Physical scheme}
\begin{figure}[h]
\includegraphics[scale=0.5]{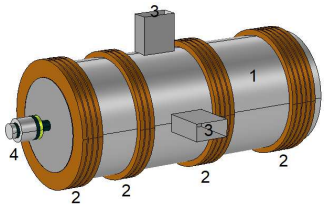}
\caption{\label{Physical_scheme} Physical model scheme: 1-cavity, 2-magnetic coils, 3-microwave ports, 4-electron gun.}
\end{figure}

Let us consider an electron injected along the axis of a cylindrical cavity excited in the Transverse-magnetic $TM_{110}$ rotating mode. The cavity is placed in an axis-symmetrical magnetostatic field produced by d.c. current coils (see Fig.\ref{Physical_scheme}).

The electron motion in autoresonant acceleration conditions is modeled by the relativistic Newton-Lorentz equation:
\begin{equation}
d\left(\gamma m_{e} \bm{v}\right) / d t=-e(\bm{E}+\bm{v} \times \bm{B}), \label{Newton-Lorentz}
\end{equation}
where $\gamma=\left[1-(v / c)^{2}\right]^{-1 / 2}$ is the Lorentz factor, $m_{e}$ and $-e$ are the mass and charge of the electron and $\bm{v}$ is its velocity. $\bm{E}=\bm{E}^{h f}$ and $\bm{B}=\bm{B}^{h f}+\bm{B}^{s}$ represent the electric and magnetic field which affect the electron motion. The superscript "$hf$" refers to the microwave field of a cylindrical rotating $TM_{110}$ mode and the superscript "$s$" refers to the axis-symmetric static magnetic field, respectively. In this manner, $\bm{B}^{s}$ indicates the field produced by magnetic coils or a permanent magnetic system. In this work, the synchrotron radiation force is not taken into consideration due to the inadequate electron energy.
\subsection{\label{sec:SaraModel}Analytical model}
\begin{figure}[h]
\includegraphics[scale=0.2]{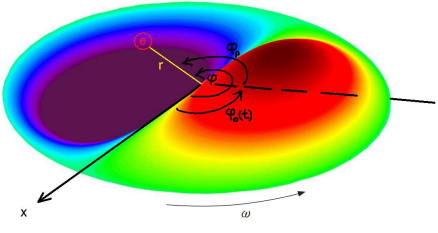}
\caption{\label{Ez_rotating} $E_z^{h f}$ component of rotating $TM_{110}$ mode and the angles used to describe the electron rotation. Here, $\varphi$ represents the angular position of the electron, $\varphi_0(t)$ defines a rotating axis where $E_z^{h f}$ has its maximum value, and $\varphi_p(t)$ is the relative angular position of the electron.}
\end{figure}

The electric and magnetic field components of the rotating TM$_{110}$ mode in the cylindrical cavity are described by the following expressions:
\begin{equation}
E_{z}^{h f}=\frac{E_{0}^{h f}}{J_{1}\left(p_{01}\right)} J_{1}\left(k_{\perp} r\right)\cos (\omega t-\varphi+\psi_0),
\end{equation}
\begin{equation}
B_{r}^{h f}=\frac{E_{0}^{h f}}{J_{1}\left(p_{01}\right)} \frac{\omega \varepsilon_0\mu_0 }{ k_{\perp}^{2}}\frac{J_{1}\left(k_{\perp} r\right)}{r}  \cos (\omega t-\varphi +\psi_0),
\end{equation}
\begin{equation}
B_{\varphi}^{h f}=\frac{E_{0}^{h f}}{J_{1}\left(p_{01}\right)} \frac{\omega \varepsilon_0 \mu_0}{k_{\perp}} J_{1}^{\prime}\left(k_{\perp} r\right) \operatorname{sin}(\omega t-\varphi+\psi_0),
\end{equation}
where $E_{0}^{h f}$ is the electric field strength of the microwave field, $J_{1}\left(k_{\perp} r\right)$ is the one-order Bessel function of first kind, $k_{\perp}=u_{11} / a$ is the radial wavenumber, being $a$ the cavity radius and $u_{1 1}=3.83171$ the first root of $J_{1}(u)$, $p_{01}=1,84118$, $\left.J_{1}^{\prime}=\left[\left(1 / k_{\perp} r\right) J_{1}\left(k_{\perp} r\right)\right]-J_{2}\left(k_{\perp} r\right)\right]$, $\omega=c k_{\perp}$ is the resonant frequency of the $TM_{110}$ mode, which does not depend of the cavity length $L$, and $\psi_0$ is an arbitrary phase.\\
From Eq.(3) we can see that $\varphi_0(t)=\omega t+\psi_0$ defines a rotating axis where $E_z^{h f}$ has its maximum value (See Fig. 2). Defining the relative angular position of the electron $\varphi_{p}(t)=\varphi(t)-\varphi_{0}(t)$, the electric and magnetic field components of the microwave field at the electron position can be written as
\begin{equation}
E_{z}^{h f}=\frac{E_{0}^{h f}}{J_{1}\left(p_{01}\right)} J_{1}\left(k_{\perp} r\right)\cos \varphi_p(t), \label{EzTM}
\end{equation} 
\begin{equation}
B_{r}^{h f}=\frac{E_{0}^{h f}}{J_{1}\left(p_{01}\right)} \frac{\omega \varepsilon_0 \mu_0 }{ k_{\perp}^{2}}\frac{J_{1}\left(k_{\perp} r\right)}{r}  \cos \varphi_p(t),
\end{equation}
\begin{equation}
B_{\varphi}^{h f}=-\frac{E_{0}^{h f}}{J_{1}\left(p_{01}\right)} \frac{\omega \varepsilon_0 \mu_0}{k_{\perp}} J_{1}^{\prime}\left(k_{\perp} r\right) \operatorname{sin}\varphi_p(t),
\end{equation}
Figure \ref{Ez_rotating} shows the local electron position and the angles which characterize the electron motion; where $\omega_{c}=d \varphi (t) / d t$ is the electron cyclotron frequency.  

To describe the magnetostatic field $\bm{B}^{s}$, the following approximation in cylindrical coordinates is used:
\begin{equation}
\bm{B}^{s}(r, z) \cong-\frac{1}{2} r \frac{d B_{z}^{s}(0, z)}{d z} \bm{\hat{r}}+B_{z}^{s}(0, z) \bm{\hat{z}} \label{Bs}
\end{equation}
where
\begin{equation}
B_{z}^{s}(0, z) \cong B_{0}\left[\gamma_{0}+b(z)\right]   \label{Bz_profile}
\end{equation}
Here, $B_{0}=m_{e} \omega / e$ is the magnetic field value corresponding to the classical cyclotron resonance, $\omega$ is the frequency of the electromagnetic field, $\gamma_{0}=[1-$ $\left.\left(v_{z 0} / c\right)^{2}\right]^{-1 / 2}$ is the Lorentz factor for the electron velocity at the injection point $z=0$, and $b(z)$ is a dimensionless function that determines the magnetic field profile along the cavity axis. This function grows in the direction of propagation of the electron and it is owing to this occurrence that the resonance conditions are maintained along the electron trajectory; therefore in order to obtain the resonance phase at the injection point, it is necessary to fulfill the prerequisite $b(0)=0$.
The magnetostatic field is presented by Expression (\ref{Bs}) in a paraxial approximation and, thus, any radial inhomogeneity can be considered to be the first-order approximation.

We focus our attention on the radial and axial components of the Lorentz force, which can be written as
\begin{equation}
F_{r}=-e\left(v_{\varphi} B_{z}^{s}-v_{z} B_{\varphi}^{h f}\right), 
\end{equation}
and
\begin{equation}
F_{z}=-e\left[E_{z}^{h f}+v_{r} B_{\varphi}^{h f}-v_{\varphi}\left(B_{r}^{h f}-\frac{1}{2} r \frac{d B_{z}^{s}}{d z}\right)\right]                     
\end{equation}
Using Eqs.(5)-(9), said components are approximate as:
\begin{equation}
\begin{aligned}
F_{r} \cong-e v_{\varphi} B_{0}\left[\gamma_{0}+b(z)\right] &-\frac{e v_{z} E_{0}^{h f} \omega \varepsilon_{0} \mu_{0}}{J_{1}\left(p_{01}\right) k_{\perp}} \\
& \times J_{1}^{\prime}\left(k_{\perp} r\right) \sin \varphi_{p}(t) \label{Fr_NL1}
\end{aligned}
\end{equation}
and
\begin{equation}
\begin{array}{r}
F_{z} \cong-\frac{e E_{0}^{h f}}{J_{1}\left(p_{01}\right)} J_{1}\left(k_{\perp} r\right) \cos \varphi_{p}(t)+e v_{\varphi} \frac{E_{0}^{h f}}{J_{1}\left(p_{01}\right)} \frac{\omega \varepsilon_{0} \mu_{0}}{k_{\perp}^{2}} \\
\times \frac{J_{1}\left(k_{\perp} r\right)}{r} \cos \varphi_{p}(t)-e v_{\varphi}\left(\frac{1}{2} r \frac{d B_{z}^{s}}{d z}\right),       \label{Fz_NL1}
\end{array}
\end{equation}
respectively. To obtain Eq. (\ref{Fz_NL1}) was taken into account that $v_rB_{\varphi}^{h f}<<E_{z}^{h f}$.\\
Equation (10) is the acting centripetal force and Eq.(11) presents the sum of the longitudinal force  exerted by both the microwave electric component (the first term), and the microwave magnetic component (the second term), and the diamagnetic force (the third term) caused by the magnetostatic field inhomogeneity, $F_{z}^{d}$. Said force is always opposite to the magnetostatic field growing, therefore it withstands the longitudinal motion of the electron. \\
To describe the electron motion, the left hand side of the Newton Lorentz equation, see Eq (\ref{Newton-Lorentz}), 
\begin{equation}
\frac{d}{d t}\left(\frac{m_e \bm{v}}{\sqrt{1-\frac{v^{2}}{c^{2}}}}\right)=\frac{m_e \bm{v}}{\left(1-\frac{v^{2}}{c^{2}}\right)^{3 / 2}}\left(\frac{\bm{v}}{c^{2}} \cdot \bm{a}\right)+\frac{m_e \bm{a}}{\sqrt{1-\frac{v^{2}}{c^{2}}}} \nonumber
\end{equation}
is written in cylindrical components as
\begin{equation}
\begin{aligned}
F_{r}=\frac{m_{e}}{\left(1-\frac{v^{2}}{c^{2}}\right)^{3 / 2}} &\left[\left(\frac{v_{r}^{2}}{c^{2}} a_{r}+\frac{v_{r} v_{\varphi} a_{\varphi}}{c^{2}}+\frac{v_{r} v_{z} a_{z}}{c^{2}}\right)\right.\\
&\left.+\left(1-\frac{v^{2}}{c^{2}}\right) a_r\right], \label{Newton_Law_Fr}
\end{aligned}
\end{equation}

\begin{equation}
\begin{gathered}
F_{\varphi}=\frac{m_{e}}{\left(1-\frac{v^{2}}{c^{2}}\right)^{3 / 2}}\left[\left(\frac{v_{\varphi} v_{r}}{c^{2}} a_{r}+\frac{v_{\varphi}^{2} a_\varphi}{c^{2}}+\frac{v_{\varphi} v_{z} a_z}{c^{2}}\right)\right. \\
\left.+\left(1-\frac{v^{2}}{c^{2}}\right) a_{\varphi}\right]
\end{gathered}
\end{equation}
and
\begin{equation}
\begin{gathered}
F_{z}=\frac{m_{e}}{\left(1-\frac{v^{2}}{c^{2}}\right)^{3 / 2}}\left[\left(\frac{v_{z} v_{r}}{c^{2}} a_{r}+\frac{v_{z} v_{\varphi}}{c^{2}}a_{\varphi}+\frac{v_{z}^{2}}{c^{2}}a_{z}\right)\right. \\
\left.+\left(1-\frac{v^{2}}{c^{2}}\right) a_{z}\right] \label{Newton_Law_Fz}
\end{gathered}
\end{equation}
where,
\begin{equation}
v_{r}=\frac{d r}{d t}, v_{\varphi}=r \frac{d \varphi}{d t}=r\omega_c(t),  v_{z}=\frac{d z}{d t},
\end{equation}
and
\begin{equation}
\begin{aligned}
&a_r=\frac{d^{2} r}{d t^{2}}-r\left(\frac{d \varphi}{d t}\right)^{2}=\frac{d^{2} r}{d t^{2}}-r \omega_{c}^{2}(t), \\
&a_{\varphi}=2\frac{ d r}{d t} \frac{d \varphi}{d t}+r \frac{d^{2} \varphi}{d t^{2}}=2 \frac{d r}{d t} \omega_{c}(t)+r \frac{d \omega_{c}}{d t}, \\
&a_{z}=\frac{d^{2} z}{d t^{2}}
\end{aligned}
\end{equation}
Because the Larmor radius does not change significantly in a cycle of the electron's helical motion, $\frac{d r}{d t}  \sim a \frac{\omega_{c}}{N}$, $\frac{d^{2} r}{d t^{2}} \sim \frac{a \omega_{c}^{2}}{N^{2}}$, being $a$ the cavity radius and $N$ the number of cycles of the electron during its motion, typically greater 10; the following conditions are fulfilled: $v_r<<c$, $a_\varphi<<a_r$ and $a_z<a_r$. Therefore, the Eqs. (\ref{Newton_Law_Fr}) and (\ref{Newton_Law_Fz}) can be approximate as:
\begin{equation}
F_{r} \cong \frac{-m_{e}}{\left(1-\frac{v^{2}}{c^{2}}\right)^{1 / 2}} a_{r}=-\gamma m_e \omega_{c}^{2} r \label{Fr_appr}
\end{equation}
and
\begin{equation}
F_{z} \cong \frac{m_{e}}{\left(1-\frac{v^{2}}{c^{2}}\right)^{3 / 2}}\left\{\frac{v_{z} v_{\varphi}}{c^{2}} \frac{d v_{\varphi}}{d t}+\left(1-\frac{v_{r}^{2}+v_{\varphi}^{2}}{c^{2}}\right) \frac{d v_{z}}{d t} \right\}. \label{Fz_appr}
\end{equation}
respectively.\\
Next, we obtain the equation describing the evolution of the relative angular position of the electron, $\varphi_p(t)$. From equations (\ref{Fr_NL1}) and (\ref{Fr_appr}) we obtain 
\begin{equation}
-\gamma m_e \omega_{c}^{2} r \cong-e v_{\varphi} B_{0}\left[\gamma_{0}+b(z)\right]-  \nonumber
\end{equation}

\begin{equation}
\frac{e
v_z E_{0}^{hf} \omega \varepsilon_{0} \mu_{0}}{J_{1}\left(p_{01}\right)k_{\perp}} J_{1}^{\prime}\left(k_{\perp} r\right) \operatorname{sin} \varphi_{p}(t)
\end{equation}
Taking into account that $v_\varphi=\omega_c r$, $\omega_0=e B_0/m_e=\omega$ (the classical cyclotron electron frequency) and $\gamma\cong\left[1- (v_\varphi^2+v_z^2)/ c^{2}\right]^{-1 / 2}$  we obtain:
\begin{equation}
\begin{array}{r}
\omega_{c}=\omega \gamma^{-1}\left\{\left[\gamma_{0}+b(\xi)\right]-u_z \left(1-\gamma^{-2}-u_z^2 \right)^{-1 / 2} \right. \\
\left.\times \left(-\frac{g_{0}^{h f}}{K_{\perp}^{*}}\right) J_{1}^{\prime}\left(k_{\perp}^* r^*\right) \sin \varphi_{p}(t)\right\} \label{omega_c}
\end{array}
\end{equation}
where $g_{0}^{hf}=E_{0}^{hf} /\left[J_1(P_{01})B_{0} c\right]$ is the dimensionless electric field strength, $u_{z}=v_{z} / c, \xi=z / r_{l}$, $k_{\perp}^*=k_{\perp} r_{l}$, $r^*=r/r_l$ and $r_{l}=c / \omega$ correspond to the relativistic Larmor radius.\\
Because $\omega_c=d \varphi(t)/dt$ and  $\varphi_{p}(t)=$ $\varphi(t)-\varphi_{0}(t)$, with $\varphi_{0}(t)=\omega t+\psi_{0}$, the time evolution of the relative angular position is described by the following expression
\begin{equation}
\dot{\varphi}_{p}=\omega_{c}(\xi) / \omega-1   \label{phidot1}
\end{equation}
where the dot denotes the operator $d/d\tau$, in which $\tau=\omega t$ (normalized time). Combining equations (\ref{omega_c}) and (\ref{phidot1}), we obtain: 
\begin{equation}
\begin{aligned}
\dot{\varphi_{p}}=\gamma^{-1}\left\{\left[\gamma_{0}-\gamma+b(\xi)\right]+\right.& \frac{g_{0}^{h f}}{k_{\perp}^{*}}u_z\left(1-\gamma^{-2}-u_{z}^{2}\right)^{-1 / 2} \\
&\left.\times J_{1}^{\prime}\left(k_{\perp}^{*} r^{*}\right) \sin \varphi_{p}\right\} \label{phidot_Bessel}
\end{aligned}
\end{equation}
Where the normalized Larmor radius can be written as 
\begin{equation}
 r^*=(1-\gamma^{-2}-u_z^2)^{1/2}/(\dot{\varphi_{p}}+1) \label{Larmor_radius1}  
\end{equation}
Taking into account that for $k_{\perp}^{*} r^{*}<1$, $J_{1}(k_{\perp}^{*} r^{*})$ is almost a linear function, we use the following approximations $J_{1}(k_{\perp}^{*} r^{*})\approx k_{\perp}^{*} r^{*}/2 $ and $J_{1}^{\prime}(k_{\perp}^{*} r^{*})\approx 1/2$; therefore the  equation (\ref{phidot_Bessel}) can be written as:
\begin{equation}
\begin{aligned}
\dot{\varphi_{p}}=\gamma^{-1}\left\{\left[\gamma_{0}-\gamma+b(\xi)\right]+\right.& \frac{g_{0}^{h f}}{2k_{\perp}^{*}}u_z\left(1-\gamma^{-2}-u_{z}^{2}\right)^{-1 / 2} \\
&\left.\times \sin \varphi_{p}\right\} \label{phidot}
\end{aligned}
\end{equation}
Replacing Eq.(\ref{phidot}) in Eqs. (\ref{phidot1}) and (\ref{Larmor_radius1}), the \textit{cyclotron electron frequency} and \textit{Larmor radius} can be written as
\begin{equation}
\omega_c=\omega \gamma^{-1}\left\{\left[\gamma_0+b(\xi)\right]+\frac{g_0^{h f}}{2 k_{\perp}^*} u_z\left(1-\gamma^{-2}-u_z^2\right)^{-1 / 2} \sin \varphi_p\right\}
\end{equation}
and
\begin{equation}
    r^*=\frac{\left(1-\gamma^{-2}-u_z^2\right) \gamma}{\left[\gamma_0+b(\xi)\right]\left(1-\gamma^{-2}-u_z^2\right)^{1 / 2}+\frac{g_0^{h f}}{2} u_z \operatorname{sen} \varphi_p } \label{Larmor_radius}
\end{equation}
respectively.\\
On the other hand, the energy evolution equation is obtained from the transferred power
\begin{equation}
\frac{d}{d t}\left(\gamma m_{e} c^{2}\right)=-e \bm{E}^{hf}\cdot\bm{v}
\end{equation}
Using Eq.(\ref{EzTM}) and the approximation for the Bessel function, we can write 
\begin{equation}
\dot{\gamma}=-g_{0}^{hf} u_{z} \left(\frac{k_{\perp}^{*} r^{*}}{2}\right) \cos \varphi_{p}, \label{gammadot_approx}
\end{equation}
Substituting Eq.(\ref{Larmor_radius}) in Eq.(\ref{gammadot_approx}) we obtain
\begin{equation}
\dot{\gamma}=-\frac{\left(\frac{g_0^{h f} k_\perp^*}{2}\right) u_z\left(1-\gamma^{-2}-u_z^2\right) \gamma \cos \varphi_p}{\left[\gamma_0+b(\xi)\right]\left(1-\gamma^{-2}-u_z^2\right)^{1 / 2}+\frac{g_0^{h f}}{2} u_z \operatorname{sen} \varphi_p} \label{gammadot}
\end{equation}

Finally, combining the equations (\ref{Fz_NL1}) and (\ref{Fz_appr}) we obtain the differential equation that describes the longitudinal motion of the electron:
\begin{equation}
\begin{gathered}
\frac{m_{e}}{\left(1-\frac{v^{2}}{c^{2}}\right)^{3 / 2}}\left\{\frac{v_{z} v_{\varphi}}{c^{2}} \frac{d v_{\varphi}}{d t}+\left(1-\frac{v_{r}^{2}+v_{\varphi}^{2}}{c^{2}}\right) \frac{d v_{z}}{d t}\right\} \cong \\
-\frac{e E_{0}^{h f}}{J_{1}\left(p_{01}\right)} J_{1}\left(k_{\perp} r\right) \cos \varphi_{p}(t)+e v_{\varphi} \frac{E_{0}^{h f}}{J_{1}\left(p_{01}\right)} \frac{\omega \varepsilon_{0} \mu_{0}}{k_{\perp}^{2}} \\
\times \frac{J_{1}\left(k_{\perp} r\right)}{r} \cos \varphi_{p}(t)-e v_{\varphi}\left(\frac{1}{2} r \frac{d B_{z}^{s}}{d z}\right),
\end{gathered}
\end{equation}
which can be written as:
\begin{equation}
\begin{aligned}
&{\left[\gamma^{-2}+\left(\frac{v_{z}}{c}\right)^{2}\right] \frac{d v_{z}}{d t} \cong-\frac{\omega}{B_{0}} \gamma^{-3}\left[1-\frac{\omega \omega_{c}(z)}{c^{2} k_{\perp}^{2}}\right] \frac{E_{0}^{h f}}{J_{1}\left(P_{01}\right)}} \\
&\times J_{1}\left(k_{\perp} r\right) \cos \varphi_{p}-\frac{v_{\varphi}^{2}}{2\left(\dot{\varphi}_{p}+1\right)} \frac{d b(z)}{d z} \gamma^{-3}-\frac{v_{z} v_{\varphi}}{c^{2}} \frac{d v_{\varphi}}{d t}. \label{long_vel}
\end{aligned}
\end{equation}
Now, from the Lorentz factor expression 
\begin{equation}
\gamma \cong\left[1-\left(\frac{v_{\varphi}^{2}+v_{z}^{2}}{c^{2}} \right) \right]^{-1/2},
\end{equation}
we have 
\begin{equation}
\frac{v_\varphi}{c^{2}} \frac{d v_{\varphi}}{d t}=\gamma^{-3} \frac{d \gamma}{d t}-\frac{v_z}{c^{2}} \frac{d v_{z}}{d t}.
\end{equation}
Therefore, Eq.(\ref{long_vel}) can be written in dimensionless units as:
\begin{equation}
\begin{gathered}
\dot{u}_{z}=\gamma^{-1}\left\{-\left(1-\frac{\dot{\varphi}_{p}+1}{k_{\perp}^{* 2}}\right) g_{0}^{h f} J_{1}\left(k_{\perp}^{*} r^{*}\right) \cos \varphi_{p}\right. \\
\left.-\frac{1-\gamma^{-2}-u_{z}^{2}}{2\left(\dot{\varphi}_{p}+1\right)} \frac{d b(\xi)}{d \xi}-u_{z} \dot{\gamma}\right\}   \label{u_z_dot_Bessel}
\end{gathered}
\end{equation}
replacing Eqs.(\ref{phidot}) and (\ref{gammadot_approx}) in Eq.(\ref{u_z_dot_Bessel}) and taking into account that $J_{1}(k_{\perp}^{*} r^{*})\approx k_{\perp}^{*} r^{*}/2$, where $r^*$ is given by Eq.(\ref{Larmor_radius}), we obtain:
\begin{equation}
\begin{aligned}
& \dot{u}_z=\left(u_z^2-1\right) g_0^{h f} \frac{k_{\perp}^*}{2} \\
& \times \frac{\left(1-\gamma^{-2}-u_z^2\right)^{1 / 2} \cos \varphi_p}{\left\{\left[\gamma_0+b(\xi)\right]+\left(\frac{g_0^{h f}}{2}\right) u_z\left(1-\gamma^{-2}-u_z^2\right)^{-1 / 2} \operatorname{sen} \varphi_p\right\}} \\
& +\quad \frac{g_0^{h f}}{2 k_{\perp}^*} \gamma^{-1}\left(1-\gamma^{-2}-u_z^2\right)^{1 / 2} \cos \varphi_p \\
& -\frac{\left(1-\gamma^{-2}-u_z^2\right)\left(\frac{d b}{d \xi}\right)}{2\left\{\left[\gamma_0+b(\xi)\right]+\left(\frac{g_0^{h f}}{2}\right) u_z\left(1-\gamma^{-2}-u_z^2\right)^{-1 / 2} \operatorname{sen} \varphi_p\right\}} \label{u_z_dot}
\end{aligned}
\end{equation}
\subsection{\label{sec:Physchem}Motion equations}
The equations (\ref{phidot}), (\ref{gammadot}) and (\ref{u_z_dot}), namely
\begin{equation}
\begin{aligned}
\dot{\varphi_{p}}=\gamma^{-1}\left\{\left[\gamma_{0}-\gamma+b(\xi)\right]+\right.& \frac{g_{0}^{h f}}{2k_{\perp}^{*}}u_z\left(1-\gamma^{-2}-u_{z}^{2}\right)^{-1 / 2} \\
&\left.\times \sin \varphi_{p}\right\} \label{phidotfinal}
\end{aligned}
\end{equation}

\begin{equation}
\dot{\gamma}=-\frac{\left(\frac{g_0^{h f} k_\perp^*}{2}\right) u_z\left(1-\gamma^{-2}-u_z^2\right) \gamma \cos \varphi_p}{\left[\gamma_0+b(\xi)\right]\left(1-\gamma^{-2}-u_z^2\right)^{1 / 2}+\frac{g_0^{h f}}{2} u_z \operatorname{sen} \varphi_p} \label{gammadot_final}
\end{equation}

\begin{equation}
\begin{aligned}
& \dot{u}_z=\left(u_z^2-1\right) g_0^{h f} \frac{k_{\perp}^*}{2} \\
& \times \frac{\left(1-\gamma^{-2}-u_z^2\right)^{1 / 2} \cos \varphi_p}{\left\{\left[\gamma_0+b(\xi)\right]+\left(\frac{g_0^{h f}}{2}\right) u_z\left(1-\gamma^{-2}-u_z^2\right)^{-1 / 2} \operatorname{sen} \varphi_p\right\}} \\
& +\quad \frac{g_0^{h f}}{2 k_{\perp}^*} \gamma^{-1}\left(1-\gamma^{-2}-u_z^2\right)^{1 / 2} \cos \varphi_p \\
& -\frac{\left(1-\gamma^{-2}-u_z^2\right)\left(\frac{d b}{d \xi}\right)}{2\left\{\left[\gamma_0+b(\xi)\right]+\left(\frac{g_0^{h f}}{2}\right) u_z\left(1-\gamma^{-2}-u_z^2\right)^{-1 / 2} \operatorname{sen} \varphi_p\right\}} \label{u_z_dot_final}
\end{aligned}
\end{equation}
and
\begin{equation}
\dot{\xi}=u_{z}, \label{zdotfinal}
\end{equation}
form a closed set of highly nonlinear differential equations, which we solve by using the fourth-order Runge Kutta method.\\
Note that at the point of injection $\xi(0)=0$, the factor $\left(1-\gamma^{-2}-\right.$ $\left.u_{z}^{2}\right)^{-1/2}$ causes a singularity. It so happens because the definition $\omega_{c}=v_{\varphi} / r$ is valid only at $r \neq 0.$ To avoid said singularity, we inject the electron at the radial distance $r_{L0}=m_e v_{0} sin \theta_0/e B_0$; where  $\theta_0$ is the small injection angle, which is measured from the cavity axis, and $v_{0}=\left(1-\gamma_{0}^{-2}\right)^{1 /2}$ is the initial velocity of the electron injected with the energy $\gamma_{0}$. Therefore, the set of equations (\ref{phidotfinal}) -(\ref{zdotfinal}) is solved by using initial conditions as follows:
\begin{equation}
\varphi_{p}\left(0\right)= \varphi_{p o},   \label{phi_0}
\end{equation}
\begin{equation}
\gamma\left(0\right)=\gamma_{0},
\end{equation}
\begin{equation}
u_{z}\left(0\right)=\left(1-\gamma_{0}^{-2}\right)^{1 / 2} \cos \theta_0, \label{uz0}
\end{equation}
and
\begin{equation}
\xi\left(0\right)=0.
\end{equation}
where $\varphi_{p o}$ is the relative angular position of the electron at the injection point, which is an arbitrary value. The choice of $\varphi_{p o}$ does not affect significantly the obtained results due to the phenomenon of phase focusing. Note that in the vicinity of the injection point, Eq.(\ref{phidotfinal}), leads to $\dot{\varphi_p} \cong \gamma^{-1} \frac{g_0^{h f}}{2 k_{\perp}^*} u_z\left(1-\gamma^{-2}-u_z^2\right)^{-1 / 2} \sin \varphi_p$. Note that if $0<\varphi_p<\pi$ then $\dot{\varphi_p}>0$; therefore the cyclotron frequency of the particle is increased, such that $\varphi_p \rightarrow \pi $. On the contrary, if $\pi<\varphi_p<2 \pi$ then $\dot{\varphi_p}<0$; therefore the electron cyclotron frequency is reduced, such that $\varphi_p \rightarrow \pi$. Due to the phase focusing, the initial phase of the electromagnetic field is irrelevant.\\ 
The rectangular coordinates of the  particle position in normalized units, $( x^*, y^*,\xi)$, are calculated as (See Fig. 2):
\begin{equation}
    x^*=r^*cos(\tau+ \varphi_p),
\end{equation}
\begin{equation}
    y^*=r^*sin(\tau+ \varphi_p),
\end{equation}
where $r^*$ is given by Eq.(\ref{Larmor_radius}), and $\xi$ is obtained from the solution of Eq.(\ref{zdotfinal}).

\section{\label{sec:Results}RESULTS AND DISCUSSIONS}

In order to check our analytical model, we compare the obtained results from the numerical solution of motion equations (\ref{phidotfinal})-(\ref{zdotfinal}) with those obtained from the numerical solution of the relativistic Newton-Lorentz equation (\ref{Newton-Lorentz}) by using the Boris method \cite{birdsall2004plasma,qin2013boris}. For the magnetostatic field profile, we consider the following three cases: (i) \textit{uniform}, (ii) \textit{linear}, and (iii) the profile to obtain \textit{exact resonance}. For all cases, we use the following simulation parameters:\\
Electric field strength: $20\mbox{ }kV/cm$,\\
Frequency: 8 GHz,\\
$\Delta t$: 0.5 ps\\
Initial conditions: $\varphi_{p0}=\pi$, $\gamma_0=1.019,1.039$ or $1.058$ (corresponding to the injection energies of $10$, $20$ and $30$ $keV$, respectively). For the condition (\ref{uz0}), we put  $\theta_0=0.01$ rad. \\
For the chosen frequency, the magnetic field for the classical resonance, $B_0$, and the cavity radius ($a=c u_{11}/2\pi f$) adopt the values of 0.286 T and 2.29 cm, respectively; while the cavity length can be arbitrarily chosen because the resonant frequency of the $TM_{110}$ mode does not depend of such value.\\

\textit{(i) Uniform magnetic field}: In this case, $B_{z}^{s}=\gamma_{0} B_0$, which corresponds to the particular case $b(\xi)=0$ and $d b(\xi)/ d\xi=0$ in equations (\ref{phidotfinal}) and (\ref{u_z_dot_final}), respectively.\\

\begin{figure}[h]
\includegraphics[scale=0.3]{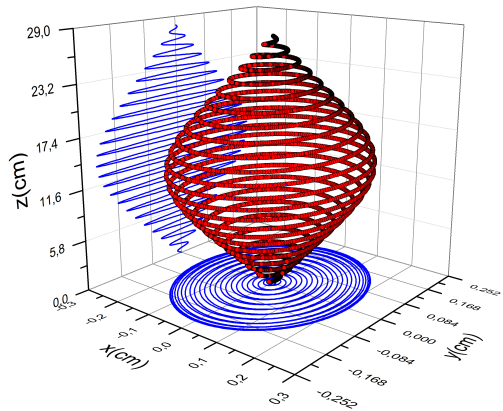}
\caption{\footnotesize The trajectory of an electron injected along the cavity axis with the energy of $30$ $keV$ for the case of the uniform magnetic field, and its projections onto the planes $xy$ and $yz$.}
\label{trajectoryBcte}
\end{figure}

\begin{figure}[h]
\includegraphics[scale=0.3]{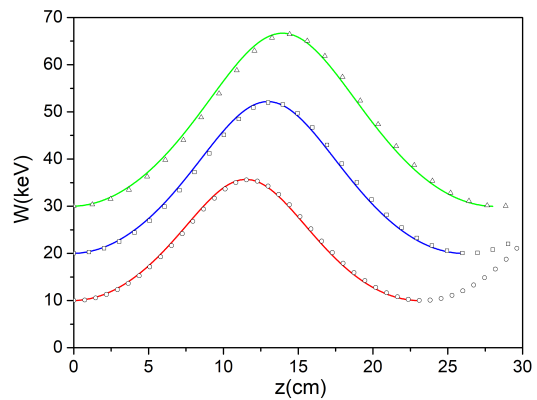}
\caption{\footnotesize Electron energy as a function of the $z$ coordinate for the injection energies of (i) 10 $keV$, (ii) 20 $keV$, and (iii) 30 $keV$, for the case of the uniform magnetostatic field of 0.286 T. The solid lines correspond to the results obtained from the model described in the present work, and the circles, squares and triangles correspond to those obtained from the numerical solution of Eq.(\ref{Newton-Lorentz}) by using the Boris method.}
\label{EnergyBcte}
\end{figure}

\begin{figure}[h]
\includegraphics[scale=0.3]{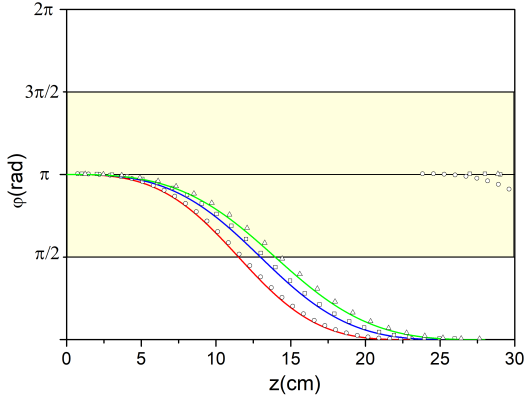}
\caption{\footnotesize  Relative angular position of the electron as a function of the $z$ coordinate for the injection energies of (i) 10 $keV$, (ii) 20 $keV$, and (iii) 30 $keV$, for the case of the uniform magnetostatic field of 0.286 T. The solid lines correspond to the results obtained from the model described in the present work, and the circles, squares and triangles correspond to those obtained from the numerical solution of Eq.(\ref{Newton-Lorentz}) by using the Boris method. The \textit{acceleration band} is highlighted in yellow color.}
\label{PhaseBcte}
\end{figure}

Figure \ref{trajectoryBcte} shows the helical trajectory of one electron injected along the cavity axis with an energy of $30$ $keV$. The electron trajectory projected in any transverse plane is a ring whose radius is a Larmor radius determined by the electron energy and the magnetic field value.\\
Figure \ref{EnergyBcte} shows the corresponding electron energies as a function of the $z$ coordinates (solid lines), which show good agreement with those obtained from the numerical solution of the relativistic Newton-Lorentz equation by using the Boris method \cite{birdsall2004plasma, qin2013boris}. For the electrons injected with the energies $10$, $20$, and $30$ keV, its energy grows monotonically up to reach the planes $z\approx 12$ cm, $z\approx 13.5$ cm, and $z \approx 14.5$ cm, respectively, and then decreases until recovering conditions similar to those of the injection point. For this reason, the system exhibits periodic behavior, not shown in Fig. \ref{EnergyBcte}, and the electron acceleration is not effective. Said results are consistent with those obtained for the relative angular position of the electron, $\varphi_p$, as a function of the $z$ coordinate shown in Fig. \ref{PhaseBcte}. Note that the electrons gain energy if $\varphi_p$ is in the range $\pi/2< \varphi_p<\pi$ and lose energy otherwise, see Eq (\ref{gammadot_approx}). For this reason, said range is named the \textit{acceleration band}. Also, in said range, the electric field component of the microwave field exerts a longitudinal force on the electron (See first right-hand side term in Eqs. (\ref{Fz_NL1}) or (\ref{u_z_dot_final})), which competes with both the longitudinal magnetic force that exerts the magnetic component of the microwave field (See second right-hand side term in Eqs. (\ref{Fz_NL1}) or (\ref{u_z_dot_final})) and the diamagnetic force which acts in the direction opposite to the magnetic field gradient and impedes the advance of the electrons into a higher magnetostatic field (See third right-hand side term in Eqs. (\ref{Fz_NL1}) or (\ref{u_z_dot_final})). For the present case $d b(\xi)/d\xi=0$.

\begin{figure}[h]
\includegraphics[scale=0.3]{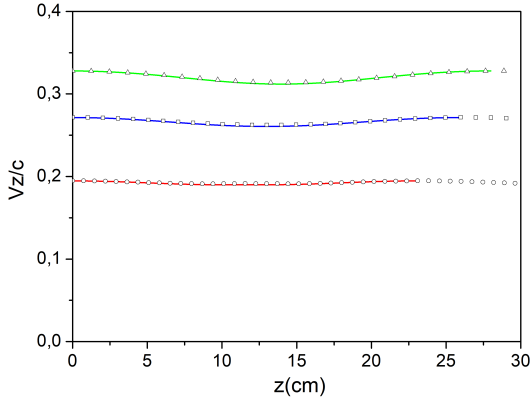}
\caption{\footnotesize Electron longitudinal velocity as a function of the $z$ coordinate for the injection energies of (i) 10 $keV$, (ii) 20 $keV$, and (iii) 30 $keV$, for the case of the uniform magnetostatic field of 0.286 T. The solid lines correspond to the results obtained from the model described in the present work, and the circles, squares and triangles correspond to those obtained from the numerical solution of Eq.(\ref{Newton-Lorentz}) by using the Boris method.}
\label{VelocityBcte}
\end{figure}

Figure \ref{VelocityBcte}, shows the electron longitudinal velocity as a function of the $z$ coordinate for the injection energies of (i) 10 $keV$, (ii) 20 $keV$, and (iii) 30 $keV$. Note that in all cases, the longitudinal velocity of the electron is nearly constant, showing that the effect of the microwave field on that velocity component is weak.\\ 

\begin{figure}[h]
\includegraphics[scale=0.3]{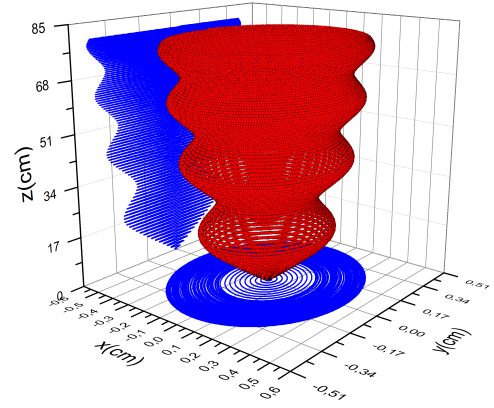}
\caption{\footnotesize The trajectory of an electron injected along the cavity axis with the energy of $30$ $keV$ for the case of the linear magnetic field.}
\label{trajectoryBlinear}
\end{figure}

\textit{(ii) Linear magnetic field}: In this case, $b(\xi)=\alpha \xi $ where $\alpha=\gamma_0 (R_m-1)/L^*$, being $L^*=L/r_l$ the cavity length in normalized units. To achieve favorable electron acceleration conditions, in the present case we use $L=0.8$ $m$ and $R_m=1.3$.\\ 
Figure \ref{trajectoryBlinear}  shows the helical trajectory of one electron injected along the cavity axis, affected by the described linear magnetostatic field, with an energy of $30$ $keV$. Note that in this case, the Larmor radius increases non-monotonically (See Fig.\ref{trajectoryBlinear} ), which allows us to infer that the energy of the electron also grows in a similar way. 

\begin{figure}[h]
\includegraphics[scale=0.3]{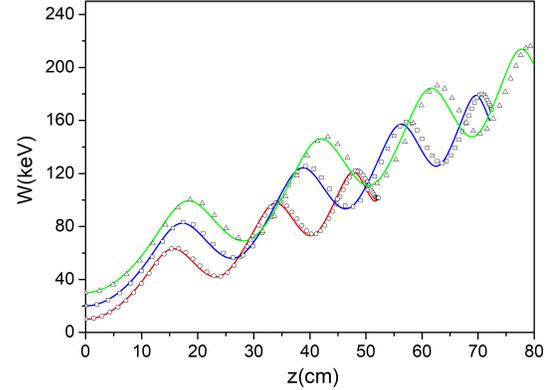}
\caption{\footnotesize Electron energy as a function of the $z$ coordinate for the injection energies of (i) 10 $keV$, (ii) 20 $keV$, and (iii) 30 $keV$, for the case of linear magnetostatic field. The solid lines correspond to the results obtained from the model described in the present work, and the circles, squares, and triangles correspond to those obtained from the numerical solution of Eq.(\ref{Newton-Lorentz}) by using the Boris method.}
\label{EnergyBlinear}
\end{figure}

Fig.\ref{EnergyBlinear} shows the electron energy as a function of the $z$ coordinate for the injection energies of (i) 10 $keV$, (ii) 20 $keV$, and (iii) 30 $keV$, for the case of the linear magnetostatic field, as a function of the $z$ coordinate. 

\begin{figure}[h]
\includegraphics[scale=0.3]{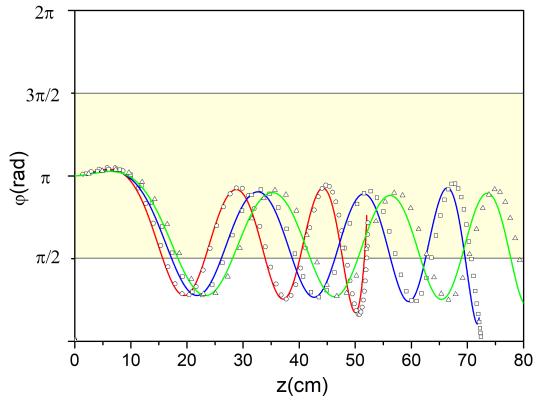}
\caption{\footnotesize  Relative angular position of the electron as a function of the $z$ coordinate for the injection energies of (i) 10 $keV$, (ii) 20 $keV$, and (iii) 30 $keV$, for the case of the linear magnetostatic field. The solid lines correspond to the results obtained from the model described in the present work, and the circles, squares, and triangles correspond to those obtained from the numerical solution of Eq.(\ref{Newton-Lorentz}) by using the Boris method.}
\label{PhaseBlinear}
\end{figure}

In this case, the magnetic field profile does allow an effective acceleration of the electron because keeps it in the acceleration band for most of its motion (See Fig.\ref{PhaseBlinear}). Note that for all cases, the spatial intervals for the $z$ coordinate where the electron is outside the acceleration band (See Fig.\ref{PhaseBlinear}) are the same intervals where the electron loses part of its energy (See Fig.\ref{EnergyBlinear}) .\\
Although the energy of the electron tends to increase with the $z$ coordinate, this tendency cannot be maintained because in this case, $d b(\xi) / d \xi \neq 0$; therefore the electron's longitudinal motion comes to a stop due to the diamagnetic force effect (See third right-hand side term in Eqs. (\ref{Fz_NL1}) or (\ref{u_z_dot_final})).

\begin{figure}[h]
\includegraphics[scale=0.3]{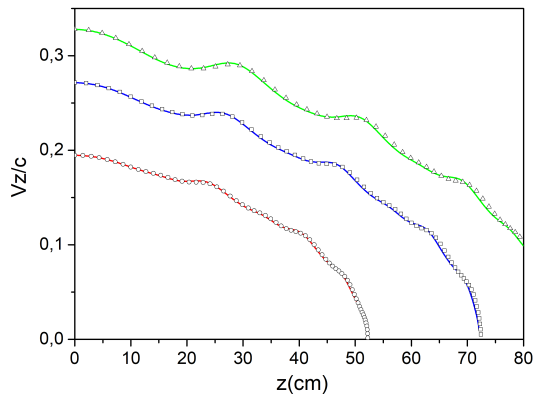}
\caption{\footnotesize Electron longitudinal velocity as a function of the $z$ coordinate for the injection energies of (i) 10 $keV$, (ii) 20 $keV$, and (iii) 30 $keV$, for the case of linear magnetostatic field. The solid lines correspond to the results obtained from the model described in the present work, and the circles, squares, and triangles correspond to those obtained from the numerical solution of Eq.(\ref{Newton-Lorentz}) by using the Boris method.}
\label{VelocityBlinear}
\end{figure}

This diamagnetic force grows as the azimuthal velocity of the electron grows as its energy increases, so the electron will inevitably stop its longitudinal motion unless the electron hits the cavity wall first. The stop plane occurs in $z=52\mbox{ }cm$ and $z=72\mbox{ }cm$ for electrons injected with the energies of $10$ $keV$ and $20$ $keV$, respectively; while the electron injected with an energy of $30$ $keV$ hits the cavity wall (See Fig.\ref{VelocityBlinear}).\\

\textit{(iii) Exact resonance magnetic field profile}: Continuous sustenance of the exact resonance can be obtained fitting the magnetostatic field profile in a way that $\varphi_{p}(\tau)=\pi$ during the electron motion. In this case, the magnetic function $b(\xi)$ is not known a priori. However,  because $\dot{\varphi_p}=0$, the Eq.(\ref{phidotfinal}) lead to:\\
\begin{equation}
    b(\xi)=\gamma-\gamma_{0} \label{b_resonance}
\end{equation}
Note that in this case, Eq. (\ref{Bz_profile}) reduces to $B_z^s(0, z) \cong \gamma B_0$; That is to say, to maintain the electron cyclotron resonance condition, the intensity of the magnetic field must increase with the $z$ coordinate at the same rate that the energy of the electron increases. For this reason, this acceleration mechanism is commonly called \textit{spatial autoresonance}.\\
Substituting Eq.(\ref{b_resonance}) in Eqs.(\ref{gammadot_final}) and (\ref{u_z_dot_final}) leads to:
\begin{equation}
\dot{\gamma}=\frac{g_0^{h f} k_\perp^*}{2} u_z\left(1-\gamma^{-2}-u_z^2\right)^{1/2} \label{gammadot_resonance}
\end{equation}
and
\begin{equation}
\dot{u}_z=-\frac{g_0^{h f} k_{\perp}^*\left(1-\gamma^{-2}-u_z^2\right)^{1 / 2}}{2 \gamma}\left\{\frac{1}{2}\left(u_z^2-1-\gamma^{-2}\right)+\frac{1}{k_{\perp}^{* 2}}\right\}  \label{uzdot_resonance},
\end{equation}
respectively. To obtain Eq.(\ref{uzdot_resonance})  it was taken into account that $d b/d \xi=u_z^{-1} \dot{\gamma}$, where $\dot{\gamma}$ is given by Eq.(\ref{gammadot_resonance}).\\Therefore, to calculate the magnetic field profile to obtain exact resonance by using Eq.(\ref{b_resonance}), first we have to solve the system of differential equations (\ref{gammadot_resonance}), (\ref{uzdot_resonance}) and (\ref{zdotfinal}), in a similar way as was described above.\\
\begin{figure}[h]
\includegraphics[scale=0.3]{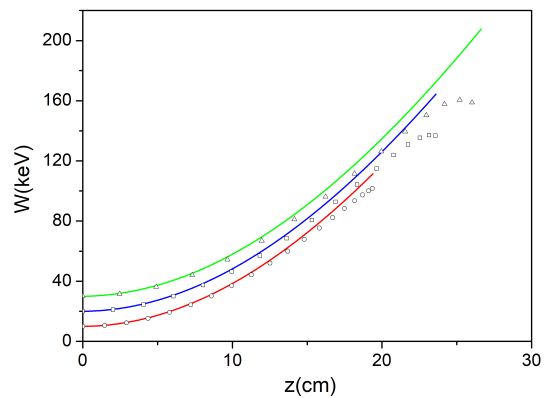}
\caption{\footnotesize Electron energy as a function of the $z$ coordinate for the injection energies of (i) 10 $keV$, (ii) 20 $keV$, and (iii) 30 $keV$, for the case of exact resonance. The solid lines correspond to the results obtained from the model described in the present work, and the circles, squares, and triangles correspond to those obtained from the Boris method.}
\label{EnergyExactResonance}
\end{figure}

\begin{figure}[h]
\includegraphics[scale=0.3]{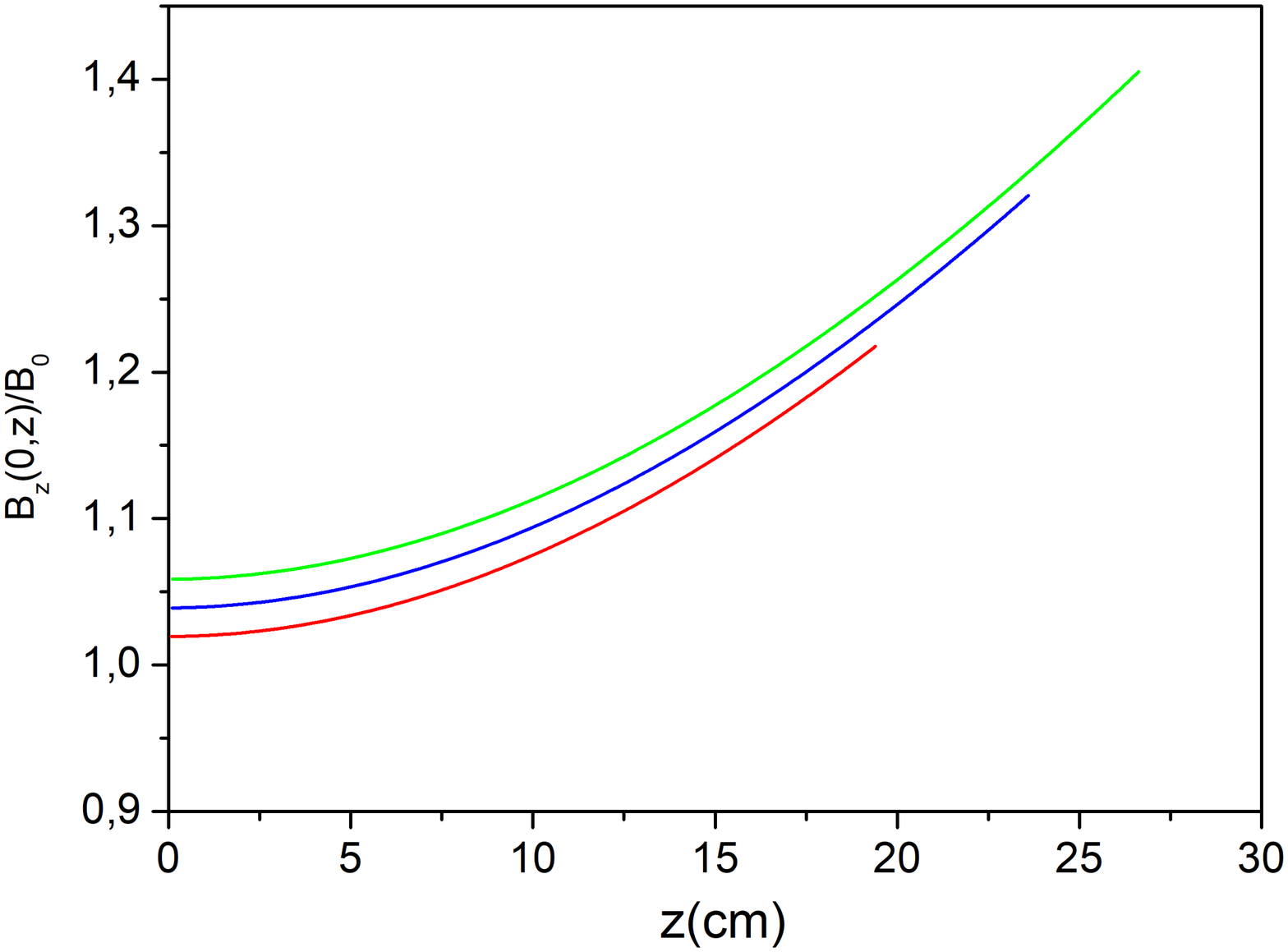}
\caption{\footnotesize Magnetic field profile to obtain exact resonance predicted by the model described in the present work for the injection energies of (i) 10 $keV$, (ii) 20 $keV$, and (iii) 30 $keV$.}
\label{Bzprofileresonance}
\end{figure}

\begin{figure}[h]
\includegraphics[scale=0.3]{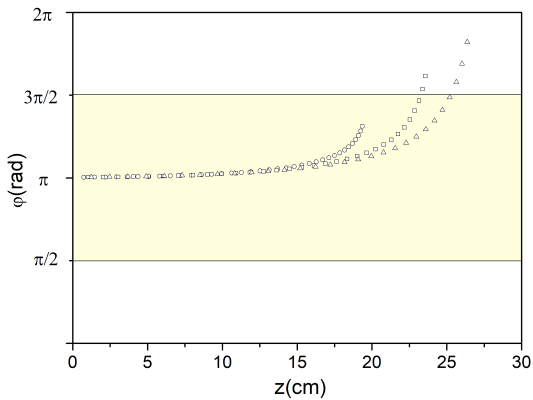}
\caption{\footnotesize  Relative angular position of the electron as a function of the $z$ coordinate for the injection energies of (i) 10 $keV$ (circles), (ii) 20 $keV$ (squares), and (iii) 30 $keV$ (triangles), obtained from the numerical solution of Eq.(\ref{Newton-Lorentz}) by using the Boris method.}
\label{PhaseBorisExactResonance}
\end{figure}

Fig.\ref{EnergyExactResonance} shows the electron energy as a function of the $z$ coordinate for the injection energies of (i) 10 $keV$, (ii) 20 $keV$, and (iii) 30 $keV$, for the case where the magnetic field profile keep the exact resonance conditions, as a function of the $z$ coordinate. Note that in the present case, as expected, the energy grows monotonically and this shows good agreement with those obtained from the numerical solution of Eq.(\ref{Newton-Lorentz}) by using the Boris method, except in the final part of said motion where the difference between these energy values is about of $15 \%$.  This so happens because the magnetic field predicted by the presented model (See Fig.\ref{Bzprofileresonance}) does not exactly correspond to the profile necessary to obtain the exact resonance condition (See Fig.\ref{PhaseBorisExactResonance}); which can be interpreted as a limitation of the analytical model.

On the other hand, for each considered injection energy, the maximum energy reached in the exact resonance condition is comparable with those obtained by using the linear magnetic field profile (See Figs.\ref{EnergyBlinear} and \ref{EnergyExactResonance}); however, in the present case, the gradient of energy is considerably larger because the electron is accelerated in maximum transferred power condition. This implies that the electron azimuthal velocity also increases faster, as well as the diamagnetic force; therefore, the electron stops its longitudinal motion more quickly. 
This effect can be appreciated in Fig.\ref{VelocityResonance}; which shows the electron longitudinal velocity as a function of the $z$ coordinate for the injection energies of (i) 10 $keV$, (ii) 20 $keV$, and (iii) 30 $keV$, for the case where the magnetic field profile keep the exact resonance conditions. Note that in the present case, the electron  stops at the $z=20$, $24$, and $27$ $cm$ planes for the three considered injection energies, respectively (See Fig.\ref{VelocityResonance}).
\begin{figure}[h]
\includegraphics[scale=0.3]{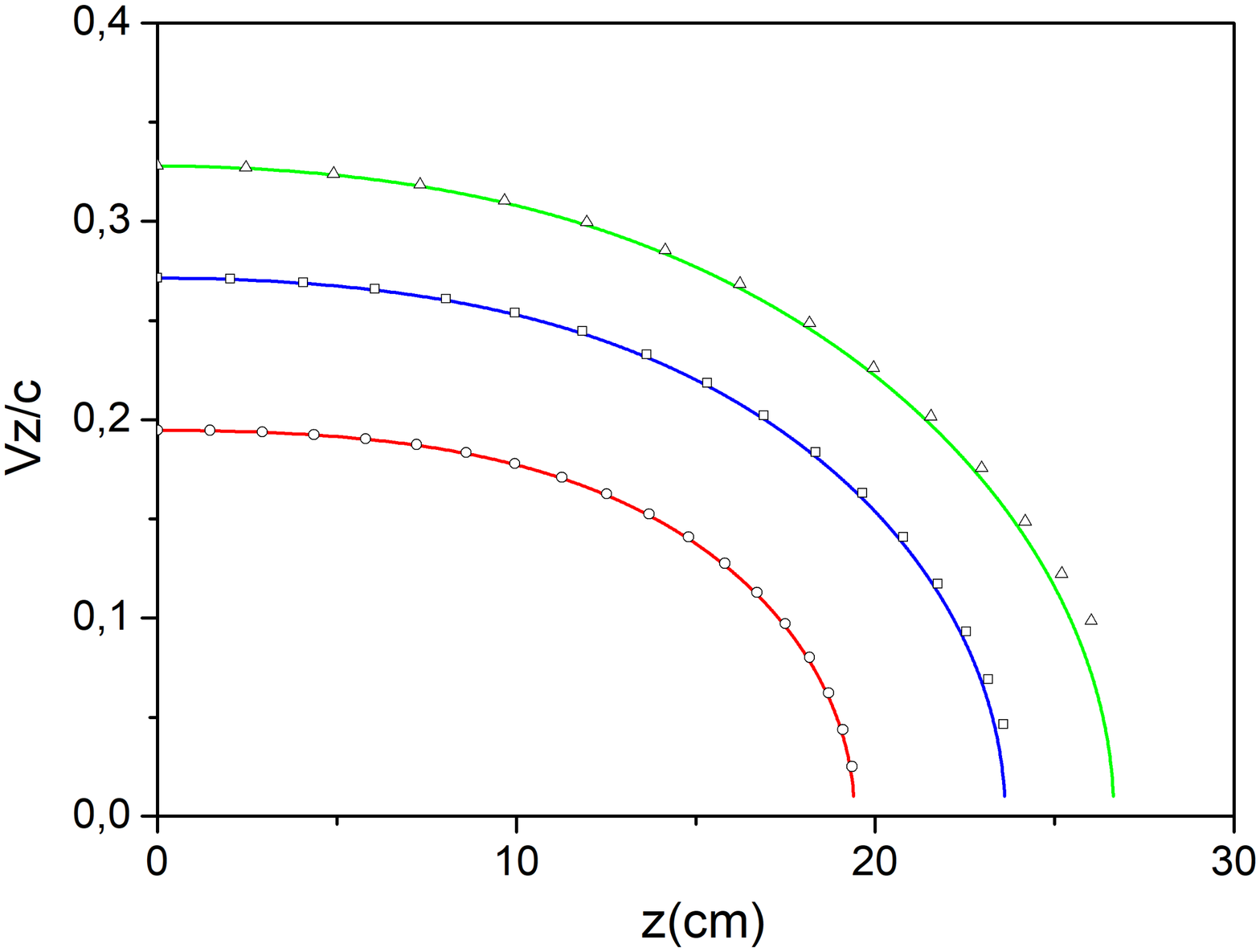}
\caption{\footnotesize Electron longitudinal velocity as a function of the $z$ coordinate for the injection energies of (i) 10 $keV$, (ii) 20 $keV$, and (iii) 30 $keV$, for the case of exact resonance. The solid lines correspond to the results obtained from the model described in the present work, and the circles, squares and triangles correspond to those obtained from the numerical solution of Eq.(\ref{Newton-Lorentz}) by using the Boris method.}
\label{VelocityResonance}
\end{figure}

\begin{figure}[h]
\includegraphics[scale=0.3]{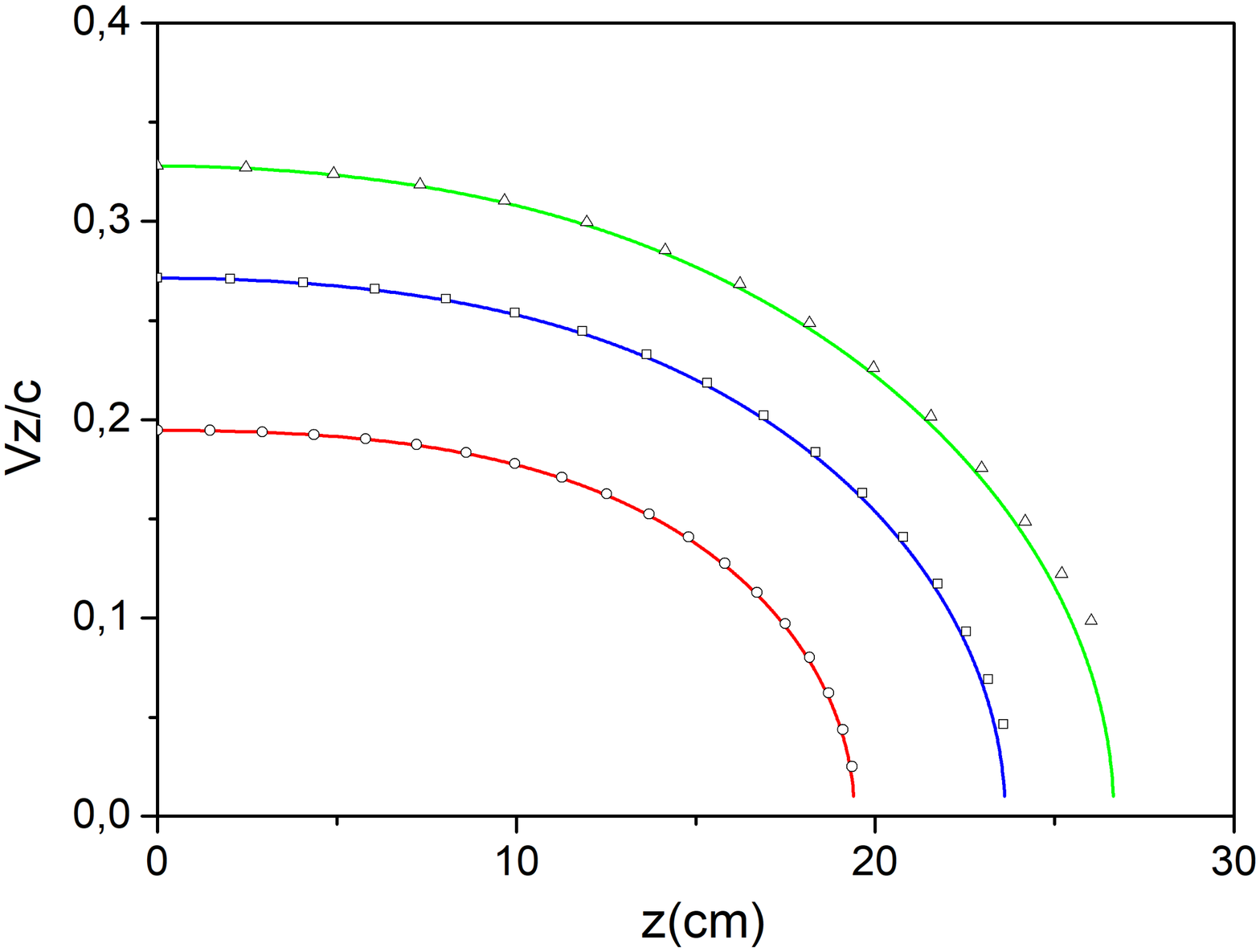}
\caption{\footnotesize Electron longitudinal velocity as a function of the $z$ coordinate for the injection energies of (i) 10 $keV$, (ii) 20 $keV$, and (iii) 30 $keV$, for the case of exact resonance. The solid lines correspond to the results obtained from the model described in the present work, and the circles, squares and triangles correspond to those obtained from the numerical solution of Eq.(\ref{Newton-Lorentz}) by using the Boris method.}
\label{VelocityResonance}
\end{figure}
\begin{figure}[h]
\includegraphics[scale=0.3]{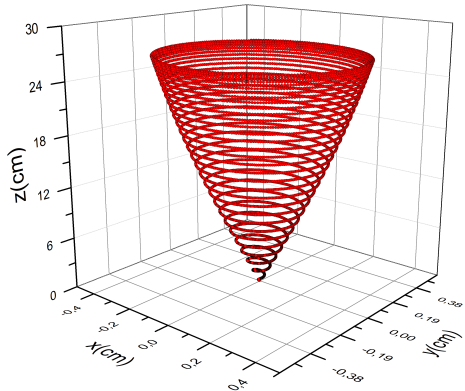}
\caption{\footnotesize The trajectory of an electron injected along the cavity axis with the energy of $30$ $keV$ for the case of exact resonance.}
\label{trajectoryResonance}
\end{figure}

Finally, Fig.\ref{trajectoryResonance} shows the helical trajectory of one electron injected with an energy of $30$ $keV$ along the cavity axis, which is affected by the magnetostatic field that guarantees the exact resonance condition. Note that in this case, unlike the cases of uniform magnetic field and linear profile magnetic field, the Larmor radius is increased monotonically due to the favorable acceleration condition for the electron.

\section{\label{sec:conclusions}CONCLUSIONS}

We have presented an analytical model for studying the cyclotron resonance acceleration of an electron by a circularly rotating $TM_{110}$ mode in an inhomogeneous magnetostatic field. Equations to calculate both the electron cyclotron frequency and the Larmor radius were obtained; which depend on the magnetic field profile, the electric field strength, the injection energy, and the motion variables. A set of differential equations describing the evolution of said variables was solved numerically. \\
Magnetic field profiles that keep the conditions for the phase shift in the acceleration band are found. We also show as the diamagnetic force, which acts in the opposite direction to the direction in which the magnetostatic field increases, imposes limits on acceleration; whose effect is more significant for the acceleration in exact resonance conditions.\\ 
The results presented in this paper can be useful for designing RF accelerators based on the circular $TM_{110}$ mode; such as  x-ray sources for medical applications or airport security. 

\begin{acknowledgments}
This work was carried out with the support of the Departamento Administrativo de Ciencia,
Tecnología e Innovación, Colombia, and Ministerio de Ciencia Tecnología e Innovación,
Colombia, through the announcement No 852-2019 (1102-852-71985) and the Universidad Industrial de Santander (UIS), Colombia, (Project ID: 9482-2665).\\
\end{acknowledgments}

\appendix

\nocite{*}

\bibliography{apssamp}

\end{document}